\documentclass[aps,twocolumn,showpacs]{revtex4}
\usepackage{graphicx}
\usepackage{amsmath}
\usepackage{amsfonts}
\usepackage{color}
\newcommand{\tbox}[1]{\mbox{\tiny #1}}

\begin{document}



\title{Dissipative fractional standard maps: Riemann-Liouville and Caputo}


\author{J. A. M\'endez-Berm\'udez}
\address{Instituto de F\'isica, Benem\'erita Universidad Aut\'onoma de Puebla, 
Puebla 72570, Mexico \\
Escuela de F\'isica, Facultad de Ciencias, Universidad Nacional Aut\'onoma 
de Honduras, Honduras}

\author{R. Aguilar-S\'anchez}
\address{Facultad de Ciencias Qu\'imicas, Benem\'erita Universidad Aut\'onoma de Puebla,
Puebla 72570, Mexico}

%

\begin{abstract}
In this study, given the inherent nature of dissipation in realistic dynamical systems,
we explore the effects of dissipation within the context of fractional dynamics.
Specifically, we consider the dissipative versions of two well known fractional maps:
the Riemann-Liouville (RL) and the Caputo (C) fractional standard maps (fSMs).
Both fSMs are two-dimensional nonlinear maps with memory given in action-angle 
variables $(I_n,\theta_n)$; $n$ being the discrete iteration time of the maps.
In the dissipative versions these fSMs are parameterized by the strength of nonlinearity $K$, 
the fractional order of the derivative $\alpha\in(1,2]$, and the dissipation strength $\gamma\in(0,1]$. 
In this work we focus on the average action $\left< I_n \right>$ and the average squared action 
$\left< I_n^2 \right>$ when~$K\gg1$, i.e.~along strongly chaotic orbits.
We first demonstrate, for $|I_0|>K$, that dissipation produces the exponential decay of the average 
action $\left< I_n \right> \approx I_0\exp(-\gamma n)$ in both dissipative fSMs.
Then, we show that while $\left< I_n^2 \right>_{\tbox{RL-fSM}}$ barely depends on $\alpha$ (effects 
are visible only when $\alpha\to 1$), any $\alpha< 2$ strongly influences the behavior of 
$\left< I_n^2 \right>_{\tbox{C-fSM}}$. We also derive an analytical expression able to describe 
$\left< I_n^2 \right>_{\tbox{RL-fSM}}(K,\alpha,\gamma)$.
\end{abstract}

\maketitle

\section{Preliminaries}

Chirikov's standard map (CSM)~\cite{C69}
\begin{equation}
\begin{array}{ll}
I_{n+1} = I_n - K\sin(\theta_n) , \\
\theta_{n+1} = \theta_n + I_{i+1} , \quad \mbox{mod}~(2\pi),
\end{array}
\label{CSM}
\end{equation}
is known to represent the local dynamics of a large family of 
Hamiltonian systems and is a paradigm model of the Kolmogorov--Arnold--Moser (KAM) scenario;
that is, the generic transition to chaos; see e.g.~Ref.~\cite{LL92}. 
The CSM, a two-dimensional nonlinear map given in action-angle 
variables, is the stroboscopic projection of the kicked rotor (KR), see e.g.~Ref.~\cite{O08}, which 
represents a free rotating stick in an inhomogeneous field that is periodically switched on in 
instantaneous pulses. The KR is described by the second order differential equation
\begin{equation}
    \frac{d^2\theta}{dt^2} + K \sin(\theta) \sum_{j = 0}^\infty \delta\left(t - j \right) = 0 .
    \label{KR}
\end{equation}
Here, $\theta\in [0,2\pi]$ is the angular position of the stick, $K$ is the kicking strength, and 
$\delta$ is the Dirac's delta function. In the CSM, $I$ corresponds to the angular momentum of the KR's stick.

By replacing the second order derivative in the equation of motion of the KR by fractional
operators (fractional derivatives, fractional integrals or fractional integro-differential operators), 
fractional versions of the KR are obtained. 
Among the many fractional KRs (fKRs) reported in the literature we can mention:
the Riemann-Liouville fKR~\cite{TZ08,ET09},
the Caputo fKR~\cite{E11}, 
the Hadamard fKR~\cite{T21d}, 
the Erdelyi-Kober fKR~\cite{T21e}, and 
the Hilfer fKR~\cite{T21f}; 
among others, see e.g.~\cite{T21b,T24}.
All the fKRs listed above, have stroboscopic versions which are known as fractional 
standard maps (fSMs), in resemblance with the CSM.

Probably, the most studied fSMs are the Riemann-Liouville fSM 
(RL-fSM)~\cite{ET09},
\begin{equation}
\begin{array}{ll}
I_{n+1} = I_n - K\sin(\theta_n) , \\
\theta_{n+1} = \displaystyle{ \frac{1}{\Gamma(\alpha)}\sum_{i = 0}^{n} I_{i+1} V^1_\alpha(n-i+1)}, \quad \mbox{mod}~(2\pi),
\end{array}
\label{RLfSM}
\end{equation}
and the Caputo fSM (C-fSM)~\cite{E11},
\begin{equation}
\begin{array}{ll}
I_{n+1} = I_n \\ 
\quad \quad \ - \displaystyle \frac{K}{\Gamma(\alpha-1)} \left[ \sum_{i = 0}^{n-1} V^2_\alpha(n-i+1)\sin(\theta_i) + \sin(\theta_n) \right]  , \\
\theta_{n+1} = \theta_n + I_0 \\
\quad \quad \ - \displaystyle \frac{K}{\Gamma(\alpha)} \sum_{i = 0}^{n} V^1_\alpha(n-i+1)\sin(\theta_i), \quad \mbox{mod}~(2\pi).
\end{array}
\label{CfSM}
\end{equation}
Here, $\Gamma$ is the Gamma function, $1<\alpha \leq 2$ is assumed, and
\begin{equation*}
	V^k_\alpha(m) = m^{\alpha-k}-(m-1)^{\alpha-k}.
\end{equation*}
Note that the sums in the equations of maps~(\ref{RLfSM}) and~(\ref{CfSM}) make the RL- fSM 
and the C-fSM to have memory, meaning that the future $(n+1)-$state depends on the entire orbit and 
not on the present $n-$state only.
Both, the RL-fSM and the C-fSM are parameterized by $K$ and $\alpha$ which control the strength of 
nonlinearity and the fractional order of the derivative, respectively.
For $\alpha=2$, both the RL-fSM and the C-fSM reproduce the CSM~\cite{C69,E11}.

It is important to stress that all three maps defined above (the CSM, the RL-fSM and the C-fSM) are 
energy-conservative maps. That is, they do not account for dissipation which is a fundamental concept 
in dynamical systems, referring to the irreversible loss of energy over time due to the interaction of a 
system with its environment, see e.g.~\cite{R06}. 

Dissipation plays a crucial role in the 
stability, behavior, and predictability of dynamical systems, particularly in real-world applications. 
In complex dynamical systems, dissipation can occur through various mechanisms such as frictional 
forces, viscous drag, turbulence, etc. These mechanisms dissipate the energy of a system, leading to 
its stabilization and eventual decay. 
The effect of dissipation on dynamical systems can be highly dependent on the nature of the system, 
its parameters, and the dissipation mechanisms involved. For example, in some systems, dissipation 
can lead to instability, while in others, it can promote stability. 

Indeed, in order to explore the effects of dissipation in generic chaotic systems, Zaslavsky introduced 
a dissipative map in Ref.~\cite{Z78}, which can also be written in the canonical form as
\begin{equation}
\begin{array}{ll}
I_{n+1} = (1-\gamma)I_n - K\sin(\theta_n) , \\
\theta_{n+1} = \theta_n + I_{i+1} , \quad \mbox{mod}~(2\pi).
\end{array}
\label{dCSM}
\end{equation}
Map~(\ref{dCSM}) is also referred to as the dissipative CSM~\cite{VZ89}.
Above, the dissipation is parametrized by $\gamma$, the dissipation strength.
If $\gamma$ equals zero in~(\ref{dCSM}), the area-preserving CSM is recovered. 
Since the determinant of the Jacobian matrix of map~(\ref{dCSM}) is $1-\gamma$, it is area-contracting 
for any $\gamma\in(0,1]$.

Furthermore, Tarasov and Edelman~\cite{TE10,T11} already introduced the dissipative version of the RL-fSM 
map as:
\begin{equation}
\begin{array}{ll}
I_{n+1} = (1-\gamma)I_n - K\sin(\theta_n) , \\
\theta_{n+1} = \displaystyle{ \frac{1}{\Gamma(\alpha)}\sum_{i = 0}^{n} I_{i+1} V^1_\alpha(n-i+1)}, \quad \mbox{mod}~(2\pi).
\end{array}
\label{dRLfSM}
\end{equation}
Moreover, in analogy, here we introduce the dissipative version of the C-fSM as
\begin{equation}
\begin{array}{ll}
I_{n+1} = (1-\gamma)I_n \\ 
\quad \quad \ - \displaystyle \frac{K}{\Gamma(\alpha-1)} \left[ \sum_{i = 0}^{n-1} V^2_\alpha(n-i+1)\sin(\theta_i) + \sin(\theta_n) \right]  , \\
\theta_{n+1} = \theta_n + I_0 \\
\quad \quad \ - \displaystyle \frac{K}{\Gamma(\alpha)} \sum_{i = 0}^{n} V^1_\alpha(n-i+1)\sin(\theta_i), \quad \mbox{mod}~(2\pi).
\end{array}
\label{dCfSM}
\end{equation}
Also note that the dissipative maps of Eqs.~(\ref{dRLfSM}) and~(\ref{dCfSM}) are, respectively, the 
Riemann-Liouville and the Caputo fractional versions of the dissipative CSM of Eq.~(\ref{dCSM}).
We recall that both dissipative fractional maps are parametrized by: the strength of nonlinearity $K$, 
the fractional order of the derivative $\alpha\in(1,2]$, and the dissipation strength $\gamma\in(0,1]$. 

Therefore, the purpose of this work is twofold. 
First, we numerically look for the effects of dissipation (parametrized by $\gamma$) in fractional 
dynamical systems, represented by the RL-fSM and the C-fSM. 
Specifically, we focus on the average action $\left< I_n \right>$ and the 
average squared action $\left< I_n^2 \right>$ when~$K\gg1$, i.e.~along strongly chaotic orbits.
Second, we obtain expressions for both $\left< I_n \right>$ and $\left< I_n^2 \right>$ which properly 
incorporates the parameter set $(K,\alpha,\gamma)$.

\section{The dissipative Riemann-Liouville fractional standard map}

We first consider the dissipative Riemann-Liouville fractional standard map (dRM-fSM) 
given in Eq.~(\ref{dRLfSM}).

\subsection{Average action $\left< I_n \right>_{\tbox{dRL-fSM}}$}
\label{Sec_avIsRL-fSM}

\begin{figure*}[ht]
\centering
\includegraphics[width=0.75\textwidth]{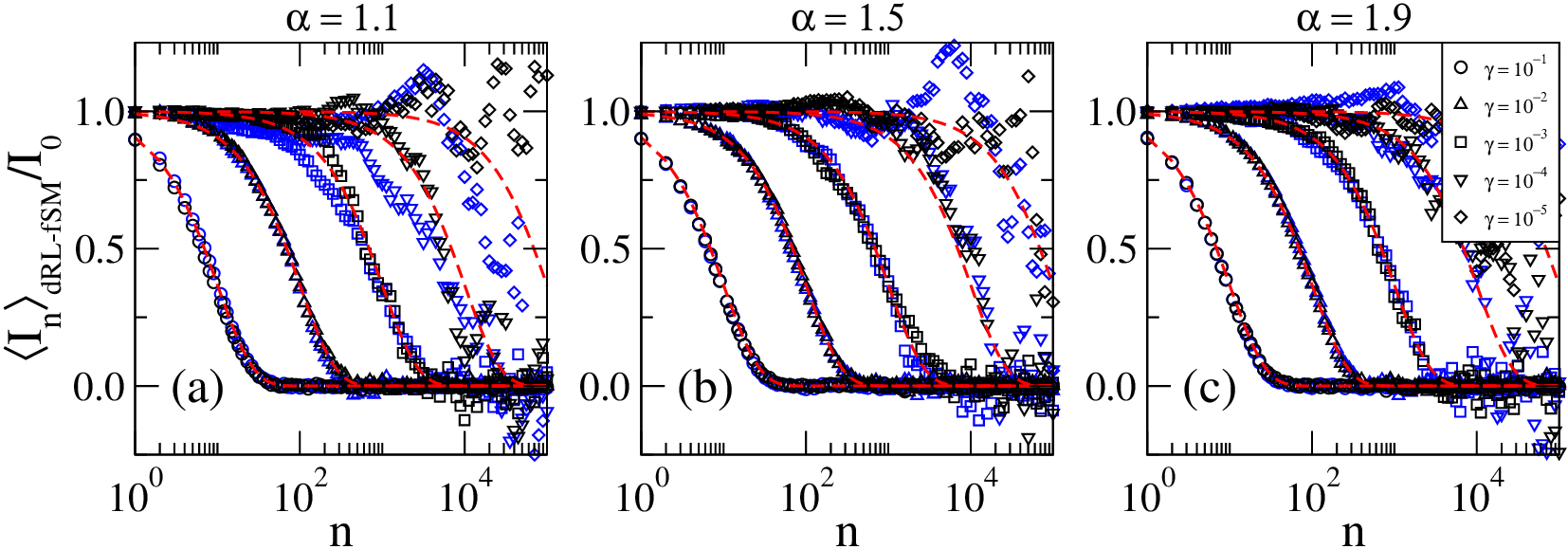}
\caption{
Average action $\left< I_n \right>$ (normalized to $I_0$) as a function of the discrete time $n$ for the 
dissipative Riemann-Liouville fractional standard map characterized by (a) $\alpha=1.1$,
(b) $\alpha=1.5$, and (c) $\alpha=1.9$. The case $I_0 > K$ is reported with $K=10^2$
(blue symbols) and $K=10^3$ (black symbols) and $I_0=10K$.
Several values of $\gamma$ are considered, as indicated in panel (c).
Red-dashed lines correspond to Eq.~(\ref{avIdRLfSM}).
The average is taken over $M=1000$ orbits with initial random phases in the interval $0<\theta_0<2\pi$.
}
\label{Fig01}
\end{figure*}

In Fig.~\ref{Fig01} we report the average action as a function of $n$ for the dRL-fSM.
We choose three representative values of $\alpha$: (a) $\alpha=1.1$,
(b) $\alpha=1.5$, and (c) $\alpha=1.9$. Here, the case $I_0 > K$ is examined with $K=10^2$
(blue symbols) and $K=10^3$ (black symbols). We set $I_0=10K$. Indeed, in order to be able to
compare curves for different values of $I_0$ we plot $\left< I_n \right>_{\tbox{dRL-fSM}}$
divided by $I_0$.
Several dissipation strengths $\gamma$ are considered, as indicated in panel (c).
The averages are computed over $M=1000$ orbits with initial random phases in the interval 
$0<\theta_0<2\pi$.

From Fig.~\ref{Fig01} we observe that $\left< I_n \right>_{\tbox{dRL-fSM}}$ decays with $n$;
the larger the dissipation strength the fastest the decay towards a chaotic attractor located
around $I \sim 0$. Indeed, the decay of $\left< I_n \right>_{\tbox{dRL-fSM}}$ is exponential
as it is demonstrated below; see also Refs.~\cite{ALM13,MOL16}.

We start with an initial condition $I_0$ and iterate the first equation of the dRL-fSM to obtain
\begin{eqnarray*}
\begin{array}{lll}
I_{1}=(1-\gamma)~I_{0}- K \sin(\theta_{0}), \\
I_{2}=(1-\gamma)^{2}~I_{0}- K [(1-\gamma) \sin(\theta_{0})+ \sin(\theta_1)], \\
I_{3}=(1-\gamma)^{3}~I_{0}- K [(1-\gamma)^{2} \sin(\theta_{0})\\
\qquad \qquad \qquad \qquad \quad + (1-\gamma) \sin(\theta_1) + \sin(\theta_{2})],\\
\quad \ \vdots \\
I_{n}=(1-\gamma)^{n}I_{0}+K\sum_{i=0}^{n-1}(1-\gamma)^{n-1-i}\sin(\theta_{i}),
\end{array}
\label{Eq2}
\end{eqnarray*}
which leads to
\begin{equation}
\left< I_n \right>_{\tbox{dRL-fSM}} = 
(1-\gamma)^{n}I_{0}+K \left< \sum_{i=0}^{n-1}(1-\gamma)^{n-1-i}\sin(\theta_{i}) \right>.
\label{Eq3}
\end{equation}
Given the periodicity of the sine function, the second term in the r.h.s. of Eq.~(\ref{Eq3}) can 
be neglected. Therefore, by expanding the first term of~(\ref{Eq3}) in powers of $n$, we obtain
\begin{eqnarray*}
\left< I_n \right>_{\tbox{dRL-fSM}} \simeq I_{0} \left[\ln(1-\gamma) n +\frac{1}{2!}
\ln(1-\gamma)^{2} n^{2} \right. \\
\left. +\frac{1}{3!} \ln(1-\gamma)^{3} n^{3}+\frac{1}{4!} \ln(1-\gamma)^{4} n^{4}+\ldots \right].
\label{Eq4}
\end{eqnarray*}
Now, considering a small value of $\gamma$ and performing a Taylor expansion we obtain that
\begin{equation}
\left< I_n \right>_{\tbox{dRL-fSM}} \simeq I_{0}e^{-\gamma n},
\label{avIdRLfSM}
\end{equation}
meaning an exponential decay towards chaotic attractors where $\gamma$ is the exponential 
decay rate.

In Fig.~\ref{Fig01} we also include Eq.~(\ref{avIdRLfSM}) as dashed lines and observe a good 
correspondence with the data, mainly for large dissipation strengths.
Note that in Fig.~\ref{Fig01} we are considering $I_0>0$ only. However, we observe the same panorama 
for $I<0$ (not shown here), once $|I_0|>K$, where Eq.~(\ref{avIdRLfSM}) is also valid.

We want to notice that the panorama reported for 
$\left< I_n \right>_{\tbox{dRL-fSM}}$ vs.~$n$ for the dRL-fSM above is equivalent to that of 
the dissipative CSM~\cite{MOL16} as well as that of the dissipative discontinuous standard 
map (DSM)~\cite{ALM13,MOL16}, both with $I_0>K\gg 1$.

We would like to stress that Eq.~(\ref{avIdRLfSM}) implies the independence of 
$\left< I_n \right>_{\tbox{dRL-fSM}}$ on $\alpha$. This, in fact, could have been
anticipated since for the derivation of Eq.~(\ref{avIdRLfSM}) we used the equation for the action in
map~(\ref{dRLfSM}), which does not explicitly depend on $\alpha$.

\subsection{Average squared action $\left< I_n^2 \right>_{\tbox{dRL-fSM}}$}
\label{Sec_avI2sRL-fSM}

\begin{figure*}[ht]
\centering
\includegraphics[width=0.75\textwidth]{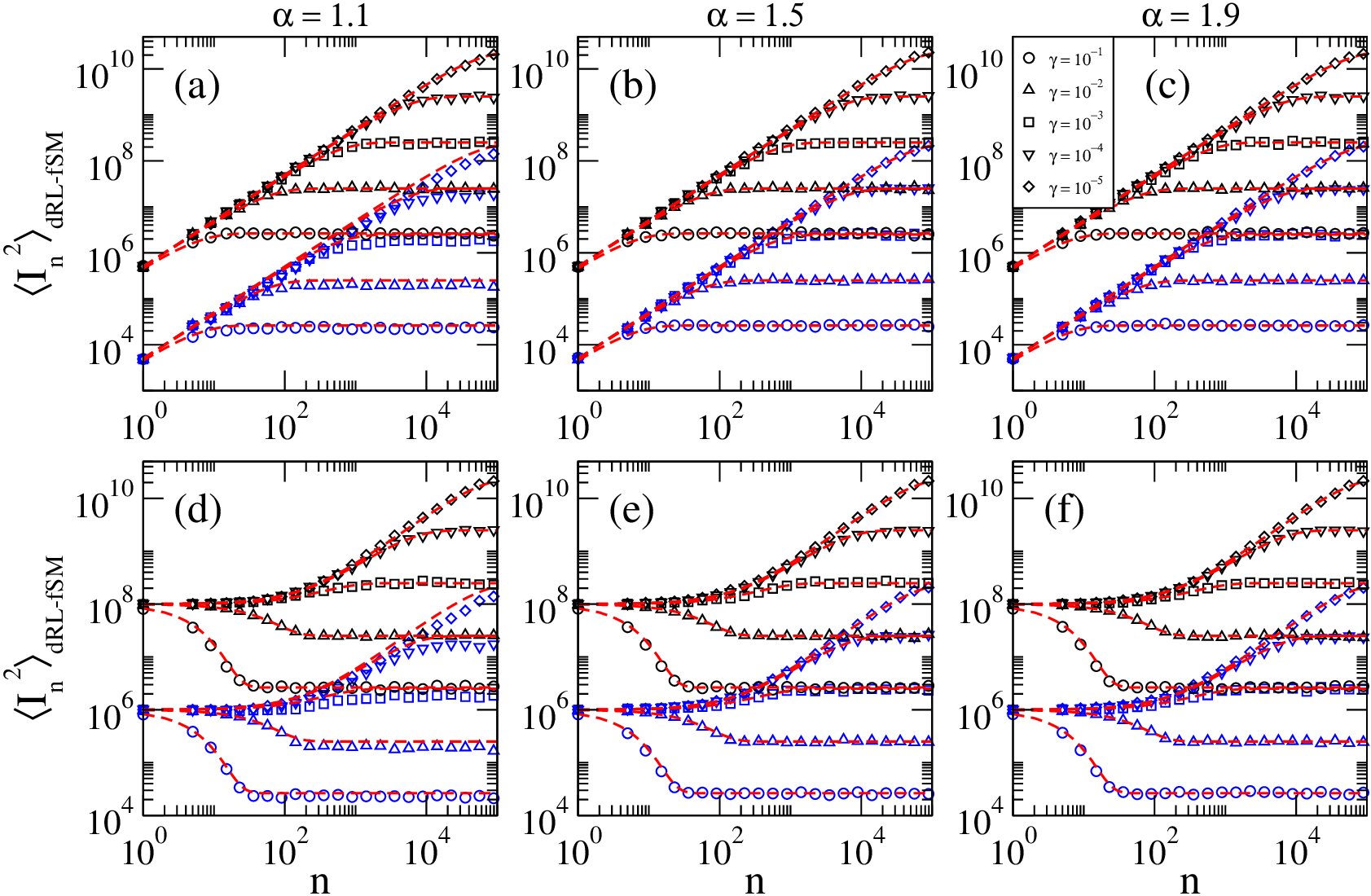}
\caption{
Average squared action $\left< I_n^2 \right>$ as a function of the discrete time $n$ for the 
dissipative Riemann-Liouville fractional standard map characterized by (a,d) $\alpha=1.1$,
(b,d) $\alpha=1.5$, and (c,f) $\alpha=1.9$. The cases (a-c) $I_0 < K$ and (d-f) $I_0 > K$ 
are reported with $K=10^2$ (blue symbols) and $K=10^3$ (black symbols). 
(a-c) $I_0=K/100$ and (d-f) $I_0=10K$ were used.
Several values of $\gamma$ are considered, as indicated in panel (c).
Red-dashed lines correspond to Eq.~(\ref{avI2dRLfSM}).
The average is taken over $M=1000$ orbits with initial random phases in the interval $0<\theta_0<2\pi$.
}
\label{Fig02}
\end{figure*}

Now, in Fig.~\ref{Fig02} we plot the average squared action as a function of $n$ for the dRL-fSM.
Again, as for $\left< I_n \right>_{\tbox{dRL-fSM}}$ above, we choose three representative values 
of $\alpha$: (a,d) $\alpha=1.1$, (b,e) $\alpha=1.5$, and (c,f) $\alpha=1.9$.
We consider $I_0<K$ and $I_0>K$ separately (Figs.~\ref{Fig02}(a-c) and Figs.~\ref{Fig02}(d-f),
respectively) since the behavior of $\left< I_n^2 \right>_{\tbox{dRL-fSM}}$ vs.~$n$
shows important differences in both cases: 
For $I_0<K$, see Figs.~\ref{Fig02}(a-c), there are two regimes separated by the crossover iteration 
time $n_{\tbox{CO}}$; a growth regime for $n<n_{\tbox{CO}}$ and the saturation regime where 
$\left< I_n^2 \right>_{\tbox{dRL-fSM}} \approx I^2_{\tbox{SAT-RL}}$ for $n>n_{\tbox{CO}}$.  
From Figs.~\ref{Fig02}(a-c) it can be observed that the nonlinearity parameter $K$
displaces the curves $\left< I_n^2 \right>_{\tbox{dRL-fSM}}$ vs.~$n$ 
vertically, while the dissipation parameter $\gamma$ determines $n_{\tbox{CO}}$. 
Moreover, there is no clear dependence of  $\left< I_n^2 \right>_{\tbox{dRL-fSM}}$ on the fractional 
order of the derivative $\alpha$. 
When $I_0>K$ the panorama is more elaborate, see Figs.~\ref{Fig02}(d-f). For small
$n$ the curves $\left< I_n^2 \right>_{\tbox{dRL-fSM}}$ vs.~$n$ are approximately
constant and equal to $I^2_0$, but there is a critical value for $\gamma$
such that if $\gamma<\gamma_{\tbox{cr}}$ or $\gamma>\gamma_{\tbox{cr}}$,
$\left< I_n^2 \right>_{\tbox{dRL-fSM}}$ increases or decreases, respectively, as a
function of $n$ before saturating at $n\to \infty$.

Since the panorama described above for $\left< I_n^2 \right>_{\tbox{dRL-fSM}}$ is equivalent to that
reported for both the CSM and the DSM~\cite{MOL16} we follow the approach reported in 
Ref.~\cite{MOL16} to get an expression for $\left< I_n^2 \right>_{\tbox{dRL-fSM}}$ vs.~$n$ as
follows (see also Ref.~\cite{MASL24}).

From the first equation of the dRL-fSM we have that
\[
I^2_{n+1}=(1-\gamma)^2I_n^2+K^2\sin^2(\theta_n)-2(1-\gamma)KI_n\sin(\theta_n),
\]
so we can write
\[
\left< I^2_{n+1} \right> = (1-\gamma)^2 \left< I_n^2 \right> + K^2 \left<\sin^2(\theta_n) \right>
+ \ 2(1-\gamma)K \left< I_n \right> \left< \sin(\theta_n) \right>.
\]
Since $\left< \sin(\theta_n) \right>=0$, the term $2(1-\gamma)K \left< I_n \right> \left< \sin(\theta_n) \right>$
can be eliminated (moreover, given the symmetry of the phase space with respect to $I=0$, 
also $\left< I_n \right>=0$). Therefore,
\begin{equation}
\label{I2b}
\left< I^2_{n+1} \right> = \left< I_n^2 \right> - (2\gamma-\gamma^2) \left< I_n^2 \right> +
\frac{K^2}{2} ,
\end{equation}
where we have already substituted $\left<\sin^2(\theta_n) \right>=1/2$.
Then, by noticing that
\begin{equation}
\label{dI2dn}
\left< I^2_{n+1} \right> - \left< I_n^2 \right> =
\frac{\left< I^2_{n+1} \right> - \left< I_n^2 \right>}{(n+1)-n}
\approx \frac{dJ}{dn} ,
\end{equation}
we rewrite Eq.~(\ref{I2b}) as the differential equation
\begin{equation}
\label{dJdn}
\frac{dJ}{dn} = -(2\gamma-\gamma^2) J + \frac{K^2}{2} ,
\end{equation}
where $J \equiv \left< I_n^2 \right>$.
Note that Eq.~(\ref{dJdn}) can be solved straightforwardly as
\begin{equation}
\int^J_{J_0} \frac{dJ'}{-(2\gamma-\gamma^2) J' + K^2/2} = \int^n_{n_0} dn' \ , \nonumber
\end{equation}
with $J_0 = \left< I_0^2 \right> = I_0^2$ and $n_0=0$.
Therefore, we finally write
\begin{equation}
\label{avI2dRLfSM}
\left< I_n^2 \right>_{\tbox{dRL-fSM}} = I_0^2e^{-(2\gamma-\gamma^2)n} +
\frac{K^2}{2(2\gamma-\gamma^2)} \left[1-e^{-(2\gamma-\gamma^2)n}\right] .
\end{equation}

Then, in Fig.~\ref{Fig02} we plot Eq.~(\ref{avI2dRLfSM}) as red-dashed lines and observe a
remarkably good correspondence with the data for both $I_0<K$ and $I_0>K$.
It is also relevant to stress that Eq.~(\ref{avI2dRLfSM}) validates the independence of
$\left< I_n^2 \right>_{\tbox{dRL-fSM}}$ on $\alpha$.
Nevertheless, it is fair to mention that the correspondence of Eq.~(\ref{avI2dRLfSM})
with the data is better the larger the value of $\alpha$ is. Notice for example that Eq.~(\ref{avI2dRLfSM})
falls above the numerical data for small $\alpha$ and small $K$; i.e.~see the blue data in 
Figs.~\ref{Fig02}(a,d).

\begin{figure*}[ht]
\centering
\includegraphics[width=0.75\textwidth]{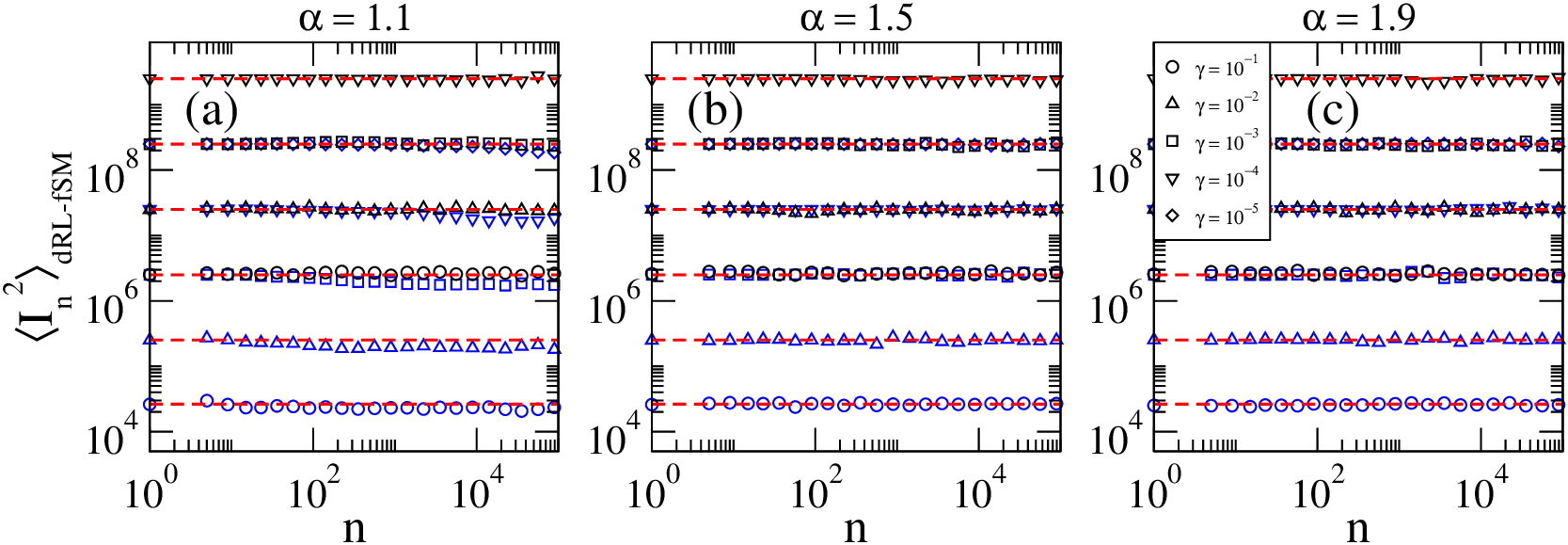}
\caption{
Average squared action $\left< I_n^2 \right>$ as a function of the discrete time $n$ for the 
dissipative Riemann-Liouville fractional standard map characterized by (a) $\alpha=1.1$,
(b) $\alpha=1.5$, and (c) $\alpha=1.9$. The case of $I_0 = I_{\tbox{SAT-RL}}$  
is reported with $K=10^2$ (blue symbols) and $K=10^3$ (black symbols).
Several values of $\gamma$ are considered, as indicated in panel (c).
Red-dashed lines correspond to Eq.~(\ref{I0}).
The average is taken over $M=1000$ orbits with initial random phases in the interval $0<\theta_0<2\pi$.
}
\label{Fig05}
\end{figure*}

In addition, Eq.~(\ref{avI2dRLfSM}) allows us to make the following observations:
\begin{itemize}

\item[\bf{(i)}]
For any $\gamma>0$ the saturation of $\left< I^2_n \right>_{\tbox{dRL-fSM}}$ is observed in the
limit of large $n$. By taking this limit into Eq.~(\ref{avI2dRLfSM}) we get
\begin{equation}
\label{I2sat}
I^2_{\tbox{SAT-RL}} \equiv \lim_{n\to\infty} \left< I^2_n \right>_{\tbox{dRL-fSM}} =
\frac{K^2}{2(2\gamma-\gamma^2)} \ .
\end{equation}

\item[\bf{(ii)}]
When $I_0<K$, case depicted in Figs.~\ref{Fig02}(a-c), the behavior
of $\left< I^2_n \right>$ as a function of $n$ is quite simple: There are two
regimes separated by the crossover iteration time $n_{\tbox{CO}}$; a growth
regime for $n<n_{\tbox{CO}}$ and the saturation regime for $n>n_{\tbox{CO}}$.
As a matter of fact, from Eq.~(\ref{avI2dRLfSM}) this situation can be well described
by
\begin{equation}
\label{I2approx}
\left< I^2_n \right>_{\tbox{dRL-fSM}} \approx I^2_{\tbox{SAT-RL}} \left[1-e^{-(2\gamma-\gamma^2)n}\right] ,
\end{equation}
which provides the growth $\left< I^2_n \right>_{\tbox{dRL-fSM}} \approx K^2 n/2$ for small $n$, 
similar to normal diffusion, and the saturation
$\left< I^2_n \right>_{\tbox{dRL-fSM}} \approx I^2_{\tbox{SAT-RL}}$ for large $n$. Moreover, from
Eq.~(\ref{I2approx}) it is clear that the ratio $\left< I^2_n \right>_{\tbox{dRL-fSM}}/I^2_{\tbox{SAT-RL}}$ 
is a {\it universal} function of the variable
$\overline{n}\equiv n/n_{\tbox{CO}}$:
\begin{equation}
\label{I2I2satscaling}
\frac{\left< I^2_n \right>_{\tbox{dRL-fSM}}}{I^2_{\tbox{SAT-RL}}} \approx 1-e^{-(2\gamma-\gamma^2)n} =
1-e^{-\overline{n}} ,
\end{equation}
where the crossover iteration time, that does not depend on $K$ nor on
$I_0$, is naturally defined as
\begin{equation}
\label{n1CO}
n_{\tbox{CO}} \equiv \frac{1}{2\gamma-\gamma^2} .
\end{equation}

\item[\bf{(iii)}]
By equating Eqs.~(\ref{avI2dRLfSM}) and (\ref{I2sat}) we assume that $\left< I^2_n
\right>$ remains constant at all times $n$, if the appropriate initial action
$I_0$ is chosen. Indeed, such $I_0$ is given by
\begin{equation}
\label{I0}
I^2_0 = \frac{K^2}{2(2\gamma-\gamma^2)} = I^2_{\tbox{SAT-RL}} .
\end{equation}
With this choice of $I_0$, dissipation and diffusion compensate each other
exactly and the squared average action does not increase nor decrease; i.e.
it remains constant and equal to $I^2_0$ (or $I^2_{\tbox{SAT-RL}}$). 
In Fig.~\ref{Fig05} we show $\left< I^2_n \right>_{\tbox{dRL-fSM}}$ as a function 
of $n$ for several combinations of $K$ and $\gamma$ (symbols). However, since we
have used as initial action the value of $I_0$ given by Eq.~(\ref{I0}), the
curves $\left< I^2_n \right>_{\tbox{dRL-fSM}}$ vs.~$n$ are close to straight horizontal lines equal to
$I^2_0$ (red full lines).
However, notice that the prediction of constant average square action is better the larger the value 
of $\alpha$ is. That is, for small $\alpha$ and small $K$, $\left< I^2_n \right>_{\tbox{dRL-fSM}}$ vs.~$n$
deviates from a straight horizontal line; i.e.~see the blue data in Fig.~\ref{Fig05}(a).

Note that given the expression for $I^2_{\tbox{SAT-RL}}$ of Eq.~(\ref{I2sat}), we can rewrite
Eq.~(\ref{avI2dRLfSM}) as
\begin{equation}
\label{I2I2satscalingB}
\left< I^2_n \right>_{\tbox{dRL-fSM}} = (I_0^2 - I^2_{\tbox{SAT-RL}}) e^{-(2\gamma-\gamma^2)n} +
I^2_{\tbox{SAT-RL}} .
\end{equation} 

\item[\bf{(iv)}]
For $I_0>K$, we recall from Figs.~\ref{Fig02}(d-f) that there is a
critical value of $\gamma$ such that if $\gamma<\gamma_{\tbox{cr}}$
or $\gamma>\gamma_{\tbox{cr}}$ then $\left< I^2_n \right>_{\tbox{dRL-fSM}}$
increases or decreases, respectively, as a function of $n$ before saturating
at $n\to \infty$. In fact, note from Eq.~(\ref{I0}) that conditions
$\gamma<\gamma_{\tbox{cr}}$ and $\gamma>\gamma_{\tbox{cr}}$ translate into
$I^2_0< I^2_{\tbox{SAT-RL}}$ and $I^2_0> I^2_{\tbox{SAT-RL}}$, respectively; therefore we obtain
\begin{equation}
\label{gammacr}
\gamma_{\tbox{cr}} = 1-\sqrt{1-\frac{K^2}{2I^2_0}} \ .
\end{equation}
For the parameters used in Figs.~\ref{Fig02}(d-f), Eq.~(\ref{gammacr}) gives 
$\gamma_{\tbox{cr}}\approx 2.5\times10^{-3}$, which coincides well with the behavior of the
data.

\end{itemize}

Finally, we want to note that the panorama reported above for the dRL-fSM  is equivalent to that of 
the dissipative CSM~\cite{MOL16} as well as for the dissipative DSM~\cite{MOL16}.
This, in fact, could have been anticipated since for the derivation of Eq.~(\ref{avI2dRLfSM}) we used the 
equation for the action in map~(\ref{dRLfSM}), which coincides with that of the dissipative CSM and the 
dissipative DSM.

\section{The dissipative Caputo fractional standard map}

We now turn our attention to the dissipative Caputo fractional standard map (dC-fSM) given in 
Eq.~(\ref{dCfSM}).

\subsection{Average action $\left< I_n \right>_{\tbox{dC-fSM}}$}

\begin{figure*}[ht]
\centering
\includegraphics[width=0.75\textwidth]{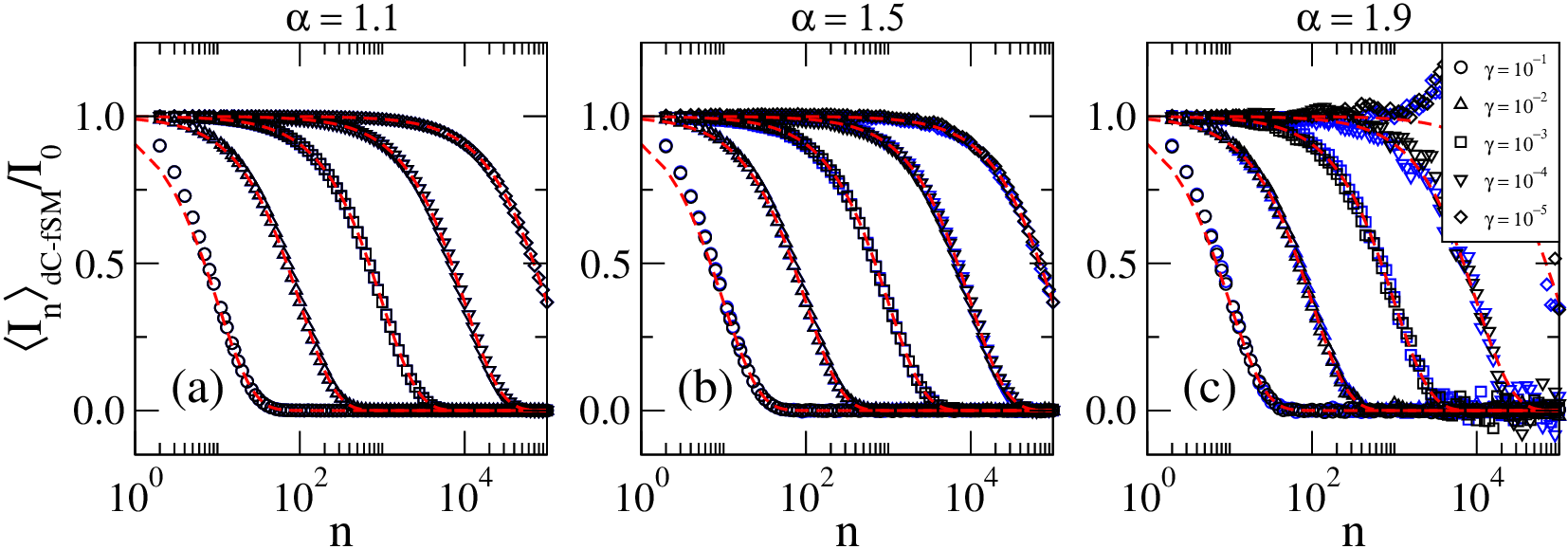}
\caption{
Average action $\left< I_n \right>$ (normalized to $I_0$) as a function of the discrete time $n$ for the 
dissipative Caputo fractional standard map characterized by (a) $\alpha=1.1$,
(b) $\alpha=1.5$, and (c) $\alpha=1.9$. The case $I_0 > K$ is reported with $K=10^2$
(blue symbols) and $K=10^3$ (black symbols) and $I_0=10K$.
Several values of $\gamma$ are considered, as indicated in panel (c).
Red-dashed lines correspond to Eq.~(\ref{avIdCfSM}).
The average is taken over $M=1000$ orbits with initial random phases in the interval $0<\theta_0<2\pi$.
}
\label{Fig03}
\end{figure*}

In Fig.~\ref{Fig03} we plot the average action as a function of $n$ for the dC-fSM.
As for the dRL-fSM, we choose three representative values of $\alpha$: (a) $\alpha=1.1$,
(b) $\alpha=1.5$, and (c) $\alpha=1.9$. In fact, for comparison purposes in Fig.~\ref{Fig03}
we use the same parameter combinations $(K,\alpha,\gamma)$ as in Fig.~\ref{Fig01} for 
the dRL-fSM.
From Fig.~\ref{Fig03} we observe that $\left< I_n \right>_{\tbox{dC-fSM}}$ decays 
with $n$ towards a chaotic attractor located around $I \sim 0$. Moreover, a
reasoning similar to the one we used in Sec.~\ref{Sec_avIsRL-fSM} to get 
Eq.~(\ref{avIdRLfSM}) lead us to
\begin{equation}
\left< I_n \right>_{\tbox{dC-fSM}} \simeq I_{0}e^{-\gamma n},
\label{avIdCfSM}
\end{equation}
which indeed reproduces well the numerical data; see the red dashed lines in Fig.~\ref{Fig03}.

It is interesting to note that the exponential decay of $\left< I_n \right>_{\tbox{dC-fSM}}$ vs.~$n$
is cleaner and emerges even at much smaller values of $\gamma$ than for 
$\left< I_n \right>_{\tbox{dRL-fSM}}$; compare Figs.~\ref{Fig01} and~\ref{Fig03}.

\begin{figure*}[ht]
\centering
\includegraphics[width=0.75\textwidth]{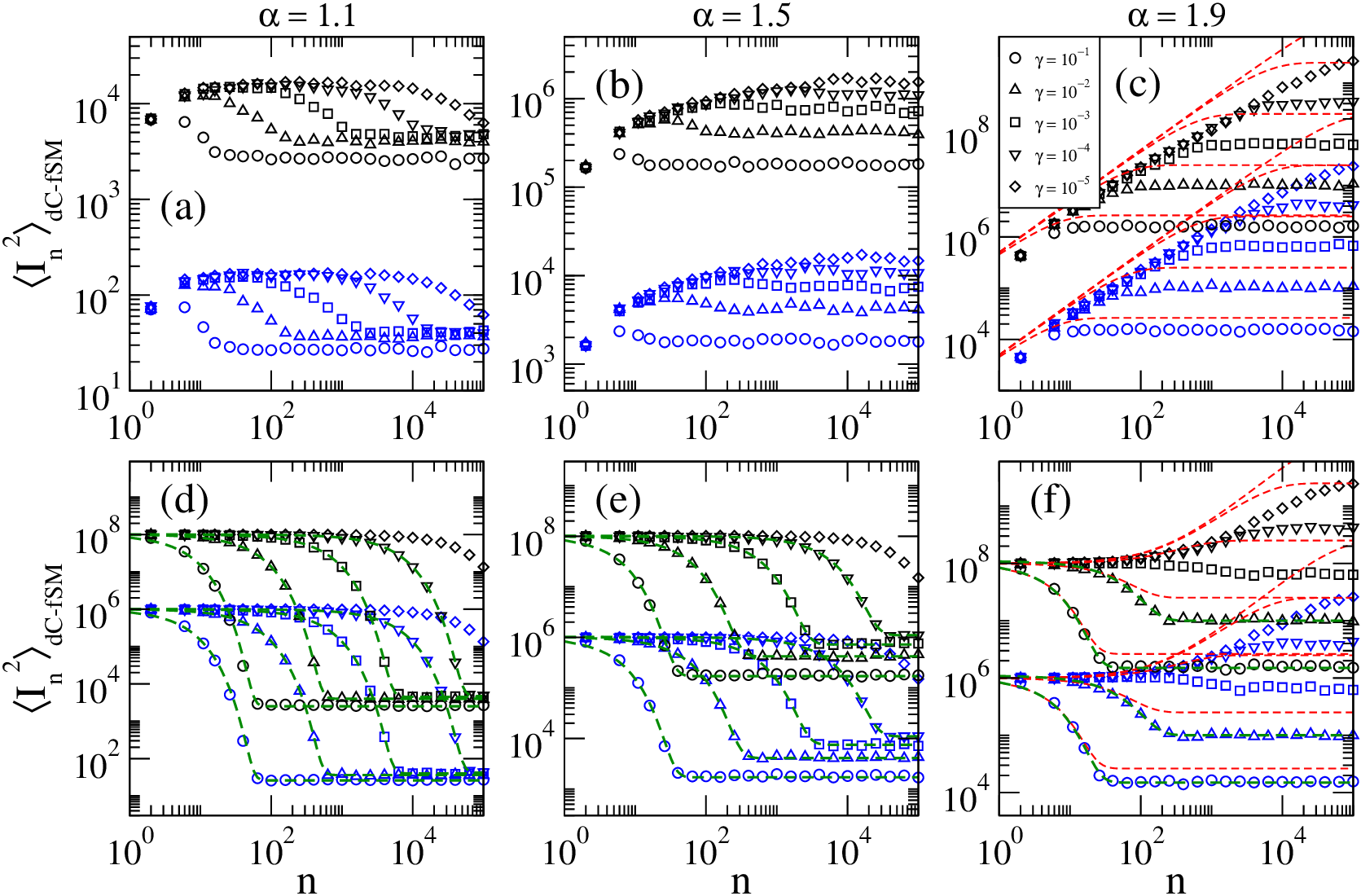}
\caption{
Average squared action $\left< I_n^2 \right>$ as a function of the discrete time $n$ for the 
dissipative Caputo fractional standard map characterized by (a,d) $\alpha=1.1$,
(b,d) $\alpha=1.5$, and (c,f) $\alpha=1.9$. The cases (a-c) $I_0 < K$ and (d-f) $I_0 > K$ 
are reported with $K=10^2$ (blue symbols) and $K=10^3$ (black symbols). 
(a-c) $I_0=K/100$ and (d-f) $I_0=10K$ were used.
Several values of $\gamma$ are considered, as indicated in panel (a).
Red-dashed lines in (c,f) correspond to Eq.~(\ref{avI2dRLfSM}).
Green-dashed lines in (d-f) correspond to Eq.~(\ref{I2I2satscalingC}).
The average is taken over $M=1000$ orbits with initial random phases in the interval $0<\theta_0<2\pi$.
}
\label{Fig04}
\end{figure*}
\begin{figure*}[ht]
\centering
\includegraphics[width=0.75\textwidth]{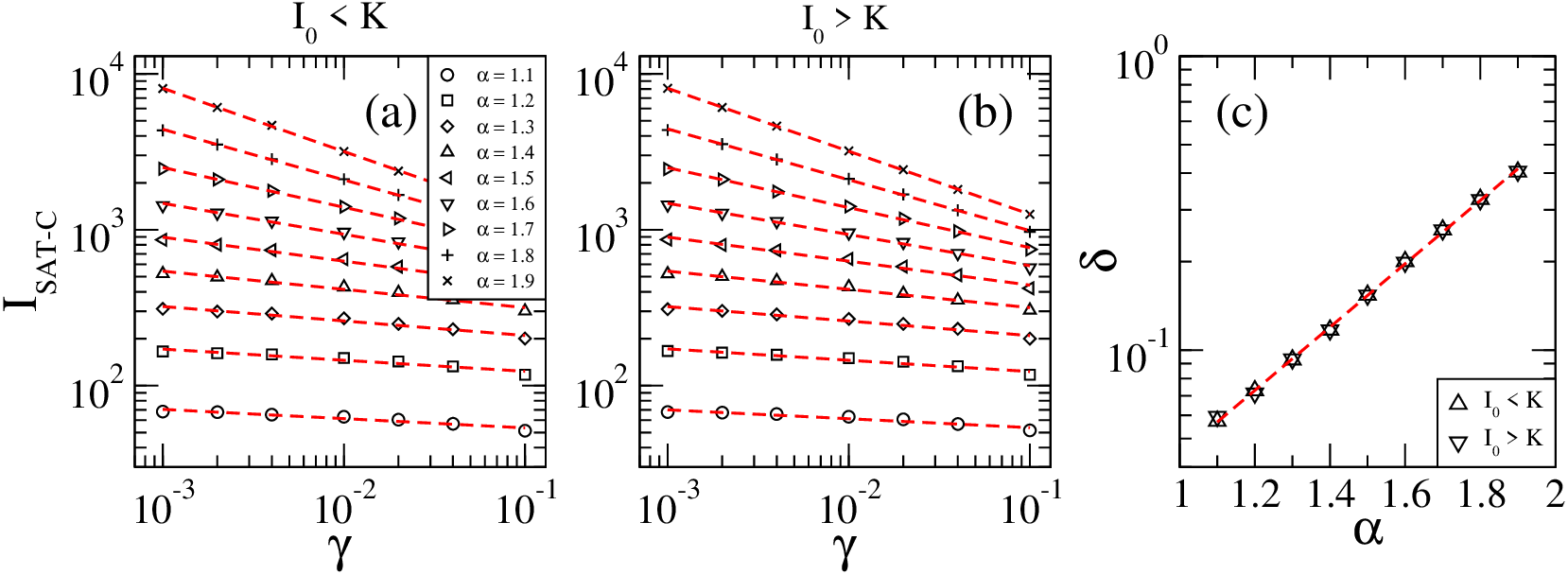}
\caption{
Average value of the action in the saturation regime $I_{\tbox{SAT-C}}$ as a function of $\gamma$ 
for the dissipative Caputo fractional standard map. 
The cases (a) $I_0 < K$ and (b) $I_0 > K$ are reported with $K=10^3$ and (a) $I_0=K/100$ and 
(b) $I_0=10K$.
Several values of $\alpha$ are considered, as indicated in panel (a).
$I_{\tbox{SAT-C}}$ is computed by averaging over $M=1000$ orbits with initial random phases in the 
interval $0<\theta_0<2\pi$.
Red-dashed lines in (a,b) correspond to fitting of Eq.~(\ref{IsatC}) to the data.
(c) Power $\delta$ as a function of $\alpha$, extracted from the power law-fittings of Eq.~(\ref{IsatC}) 
to the data in panels (a,b). 
Red-dashed line in (c) is an exponential fitting to the data leading to $\delta \approx 0.0034\exp (5\alpha/2)$.
}
\label{Fig06}
\end{figure*}

\subsection{Average squared action $\left< I_n^2 \right>_{\tbox{dC-fSM}}$}

Now, in Fig.~\ref{Fig04} we pesent the average squared action as a function of $n$ for the dC-fSM.
We again use the three values of $\alpha$: (a,d) $\alpha=1.1$, (b,e) $\alpha=1.5$, and (c,f) $\alpha=1.9$. 
For comparison purposes in Fig.~\ref{Fig04} we use the same parameter combinations 
$(K,\alpha,\gamma)$ as in Fig.~\ref{Fig02} for the dRL-fSM.

From Fig.~\ref{Fig04} we can clearly see that $\left< I^2_n \right>_{\tbox{dC-fSM}}$ is strongly
affected by the fractional order of the derivative $\alpha$; this in contrast with 
$\left< I^2_n \right>_{\tbox{dRL-fSM}}$, which is independent of $\alpha$.
Indeed, the smaller the value of $\alpha$ the stronger the deviation between 
$\left< I^2_n \right>_{\tbox{dC-fSM}}$ and Eq.~(\ref{avI2dRLfSM}), which describes 
$\left< I^2_n \right>_{\tbox{dRL-fSM}}$ and is included as red-dashed lines in Figs.~\ref{Fig04}(c,f)
as a reference.
In fact, while for $\alpha=1.9$ the curves of $\left< I^2_n \right>_{\tbox{dC-fSM}}$ vs.~$n$
are relatively close to Eq.~(\ref{avI2dRLfSM}) (as expected since both dissipative maps, the 
dRL-fSM and the dC-fSM, coincide for $\alpha\to 2$), see Figs.~\ref{Fig04}(c,f); for $\alpha=1.1$
the behavior of $\left< I^2_n \right>_{\tbox{dC-fSM}}$ is very different to that of 
$\left< I^2_n \right>_{\tbox{dRL-fSM}}$, see Figs.~\ref{Fig04}(a,d):
For example, for $I_0>K$, $\left< I^2_n \right>_{\tbox{dC-fSM}}$ decays with $n$ for all $\gamma$,
see Fig.~\ref{Fig04}(d).

It is now important to admit that, unfortunately, by following the simple arguments used in 
Sec.~\ref{Sec_avI2sRL-fSM} to get Eq.~(\ref{avI2dRLfSM}), we are not able to get an expression 
for $\left< I^2_n \right>_{\tbox{dC-fSM}}$. 
However, we have observed that for $I_0>K$, the expression
\begin{equation}
\label{I2I2satscalingC}
\left< I^2_n \right>_{\tbox{dC-fSM}} \approx I_0^2 e^{-(2\gamma-\gamma^2)n} +
I^2_{\tbox{SAT-C}} 
\end{equation} 
which is equivalent to Eq.~(\ref{I2I2satscalingB}), describes well the data when $\alpha\to 1$;
see the green dashed lines in Figs.~\ref{Fig04}(d-f).

We have also observed that $I_{\tbox{SAT-C}}$ strongly depends on $\alpha$,
so it deserves a careful characterization.
First, we noticed that $I_{\tbox{SAT-C}}\propto K$, as well as $I_{\tbox{SAT-RL}}$; see Eq.~(\ref{I2sat}).
Then, in Figs.~\ref{Fig06}(a,b) we plot $I_{\tbox{SAT-C}}$ as a function of $\gamma$ 
for the dC-fSM with (a) $I_0<K$ and (b) $I_0>K$ for several fixed values of $\alpha$, as 
indicated in panel (a).
From Figs.~\ref{Fig06}(a,b) we can clearly see the independence of $I_{\tbox{SAT-C}}$ on $I_0$
and the power-law dependence
\begin{equation}
I_{\tbox{SAT-C}} \propto \gamma^\delta , \quad \delta\equiv\delta(\alpha) .
\label{IsatC}
\end{equation}

Moreover, to look for the dependence of $\gamma$ on the fractional order of the Caputo derivative,
in Figs.~\ref{Fig06}(a,b) we perform fitting of Eq.~(\ref{IsatC}) to the data, see the red-dashed lines.
Therefore, we report the obtained values of $\gamma$ in Fig.~\ref{Fig06}(c).
From Fig.~\ref{Fig06}(c) we can conclude that
$\delta$ depends exponentially on $\alpha$, specifically we get $\delta \approx 0.0034\exp (5\alpha/2)$;
see the red dashed line.
With this we also confirm that $\delta\to 1/2$ when $\alpha\to 2$ as expected from Eq.~(\ref{I2sat}).

\begin{figure*}[ht]
\centering
\includegraphics[width=0.75\textwidth]{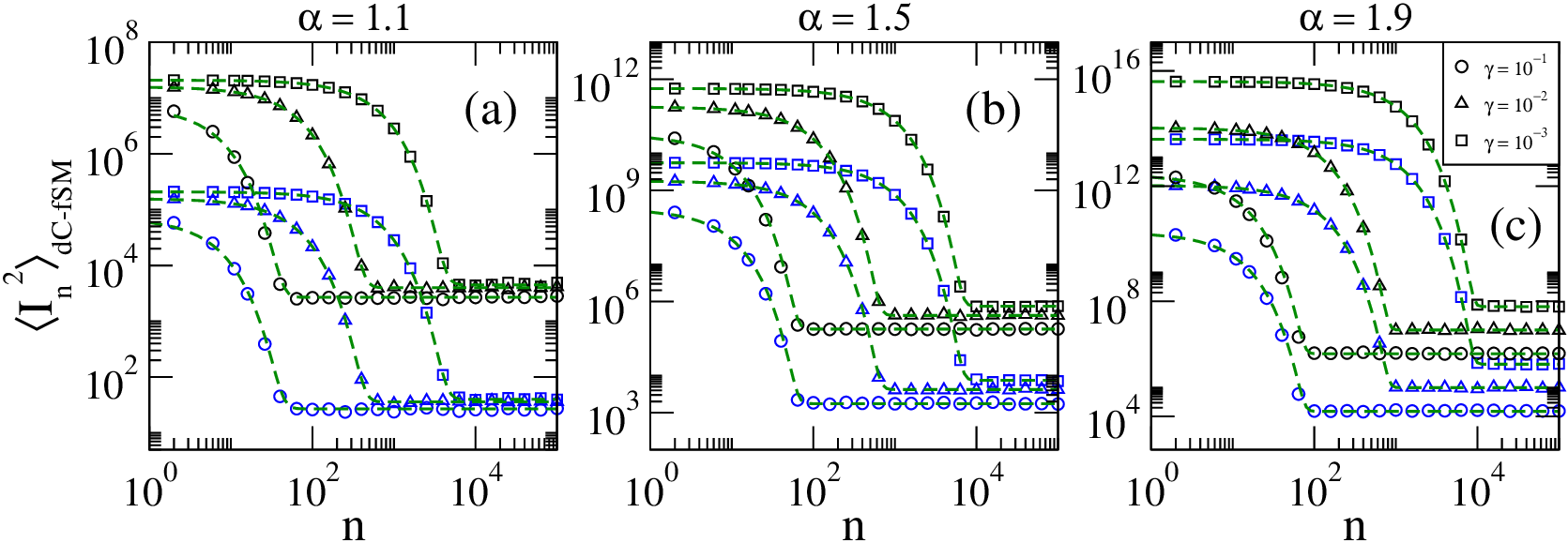}
\caption{
Average squared action $\left< I_n^2 \right>$ as a function of the discrete time $n$ for the 
dissipative Caputo fractional standard map characterized by (a) $\alpha=1.1$,
(b) $\alpha=1.5$, and (c) $\alpha=1.9$. The case of $I_0 \approx I_{\tbox{SAT-C}}$  
is reported with $K=10^2$ (blue symbols) and $K=10^3$ (black symbols).
Several values of $\gamma$ are considered, as indicated in panel (c).
Green-dashed lines correspond to Eq.~(\ref{I2I2satscalingC}).
The average is taken over $M=1000$ orbits with initial random phases in the interval $0<\theta_0<2\pi$.
}
\label{Fig07}
\end{figure*}

\section{Discussion and conclusions}

In this work we have considered the dissipative versions of two well-known fractional maps:
the dissipative Riemann-Liouville fractional standard map (dRL-FSM) and 
the (introduced here) dissipative Caputo fractional standard map (dC-FSM). So
we explored the effects of dissipation within the context of fractional dynamics.
Both, the dRL-fSM and the dC-fSM are parameterized by the strength of nonlinearity $K>0$, 
the fractional order of the derivative $\alpha\in(1,2]$, and the dissipation strength $\gamma\in(0,1]$.
Specifically, we focused on the average action $\left< I_n \right>$ and the average squared action 
$\left< I_n^2 \right>$ when~$K\gg1$, i.e.~along strongly chaotic orbits.

First, we demonstrated, for $|I_0|>K$, that dissipation produces the exponential decay of the average 
action in both dissipative fSMs, see Eqs.~(\ref{avIdRLfSM}) and~(\ref{avIdCfSM}) as well as 
Figs.~\ref{Fig01} and~\ref{Fig03}.

Second, we showed that $\left< I_n^2 \right>_{\tbox{RL-fSM}}$ barely depends on $\alpha$, see
Fig.~\ref{Fig02}, and derived an analytical expression able to properly describe 
$\left< I_n^2 \right>_{\tbox{RL-fSM}}(K,\alpha,\gamma)$, see Eq.~(\ref{avI2dRLfSM}).
Also we observed for $n\to\infty$ that $ \left< I^2_n \right>_{\tbox{dRL-fSM}}$ converges to a saturation 
value, $I^2_{\tbox{SAT-RL}}$,  and found an expression for it, see Eq.~(\ref{I2sat}).
Moreover, from Eq.~(\ref{I2I2satscalingB}), which is written by substituting Eq.~(\ref{I2sat}) into 
Eq.~(\ref{avI2dRLfSM}), we can recognize
the following {\it universal} function of the variable $\overline{n}$:
\begin{equation}
\label{I2I02scaling3}
\overline{\left< I^2_n \right>}_{\tbox{dRL-fSM}} \equiv \frac{\left< I^2_n \right>_{\tbox{dRL-fSM}} - I^2_{\tbox{SAT-RL}}}{I_0^2  - I^2_{\tbox{SAT-RL}}} \approx
e^{-\overline{n}} \ ,
\end{equation}
which embraces all the scenarios reported in Fig.~\ref{Fig02}.
Here $\overline{n}\equiv n/n_{\tbox{CO}}$ where $n_{\tbox{CO}}$ is given by Eq.~(\ref{n1CO}).

Third, we showed that any $\alpha< 2$ strongly influences the behavior of $\left< I_n^2 \right>_{\tbox{dC-fSM}}$,
see Fig.~\ref{Fig04}, and obtained a phenomenological expressions for $\left< I_n^2 \right>_{\tbox{dC-fSM}}$, 
see Eq.~(\ref{I2I2satscalingC}), which works in certain parameter regimes.
Moreover, in contrast with $I^2_{\tbox{SAT-RL}}$, we found that the saturation value for the average
squared action of the dC-fSM strongly depends on $\alpha$, see Eq.~(\ref{IsatC}) and Figs.~\ref{Fig04}
and~\ref{Fig06}.

Finally, it is interesting to highlight that from the analysis of Eq.~(\ref{avI2dRLfSM}) we were able to
identify the condition, i.e.~Eq.~(\ref{I0}), for which dissipation and diffusion compensate each other 
exactly so the squared average action of the dRL-fSM remains constant. That is, by setting $I_0 = I_{\tbox{SAT-RL}}$,
the curves $\left< I_n^2 \right>_{\tbox{dRL-fSM}}$ vs.~$n$ are straight horizontal lines, except for
$\alpha\to 1$ where $\left< I_n^2 \right>_{\tbox{dRL-fSM}}$ slightly decreases with $n$; see 
Fig.~\ref{Fig05}.
A pertinent question now is whether we can observe a similar behavior for 
$\left< I_n^2 \right>_{\tbox{dC-fSM}}$. Thus, in Fig.~\ref{Fig07} we report $\left< I_n^2 \right>_{\tbox{dC-fSM}}$
vs.~$n$, for several combinations of $\gamma$ and $\alpha$, where we have used $I_0 \approx I_{\tbox{SAT-C}}$, 
where $ I_{\tbox{SAT-C}}$ has been computed numerically; see e.g.~Fig.~\ref{Fig06}.
It is clear from Fig.~\ref{Fig07} that $\left< I_n^2 \right>_{\tbox{dC-fSM}}$ does not remain constant as
a function of time, instead it first decreases exponentially and approaches a {\it second} saturation value.
Indeed, Eq.~(\ref{I2I2satscalingC}) reproduces well the numerical data, see the green-dashed lines.
The reason for this difference between the dRL-fSM and the dC-fSM is the absence/presence of the 
sum (which implies memory) in the equation for the action in the dRL-fSM/dC-fSM. So, to maintain
the saturation value in the dC-fSM the entire orbit is needed.

We hope our results may stimulate further numerical as well as analytical studies on 
dissipative fractional dynamics.

\section*{Acknowledgements}

J.A.M.-B. thanks support from VIEP-BUAP (Grant No.~100405811-VIEP2024), Mexico.


\begin{thebibliography}{00}

\bibitem{C69}
B. V. Chirikov, 
Research concerning the theory of nonlinear resonance and stochasticity,
Preprint 267, Institute of Nuclear Physics, Novosibirsk (1969). Engl. Trans., CERN Trans. (1971) 71-40.

\bibitem{LL92}
A. J. Lichtenberg and M. A. Lieberman, 
Regular and Chaotic Dynamics (Springer-Verlag, New York, 1992).

\bibitem{O08}
E. Ott, 
Chaos in dynamical systems
(Cambridge Univ. Press, 2008).

\bibitem{TZ08}
V. E. Tarasov and G. M. Zaslavsky, 
Fractional equations of kicked systems and discrete maps,
J. Phys. A {\bf 41} (2008) 435101.

\bibitem{ET09} 
M. Edelman and V. E. Tarasov,
Fractional standard map,
Phys. Lett. A {\bf 374} (2009) 279--285.

\bibitem{E11}
M. Edelman,
Fractional standard map: Riemann-Liouville vs. Caputo, 
Commun. Nonlinear. Sci. numer. Simulat. {\bf 16} (2011)
4573--4580.

\bibitem{T21d}
V. E. Tarasov,
Fractional dynamics with non-local scaling,
Commun. Nonlinear Sci. Numer. Simulat. {\bf 102} (2021) 105947.

\bibitem{T21e}
V. E. Tarasov,
Nonlinear fractional dynamics with kicks,
Chaos Solitons Fractals {\bf 151} (2021) 111259.

\bibitem{T21f}
V. E. Tarasov,
From fractional differential equations with Hilfer derivatives to discrete maps with memory,
Com. Appl. Math. {\bf 40} (2021) 296.

\bibitem{T21b}
V. E. Tarasov,
Integral equations of non-integer orders and discrete maps with memory,
Mathematics {\bf 9} (2021) 1177.

\bibitem{T24}
V. E. Tarasov,
Discrete maps with distributed memory fading parameter,
Comput. Appl. Math. {\bf 43} (2024) 113.

\bibitem{R06}
M. Razavy,
Classical and Quantum Dissipative Systems
(World Scientific, 2006). 

\bibitem{Z78} 
G. M. Zaslavsky,
The simplest case of a strange attractor,
Phys. Lett. A {\bf 69} (1978) 145--147. 

\bibitem{VZ89} 
O. F. Vlasova and G. M. Zaslavsky
Nonergonic regions in the standard dissipative mapping
Phys. Lett. A {\bf 105} (1984) 1--5.

\bibitem{TE10} 
V. E. Tarasov, and M. Edelman,
Fractional dissipative standard map,
Chaos {\bf 20} (2010) 023127.

\bibitem{T11} 
V. E. Tarasov,
\textit{Fractional Zaslavsky and Henon discrete maps},
Chapter 1 in Long-range Interaction, Stochasticity and Fractional Dynamics, 
A. C. J. Luo and V. Afraimovich (Eds.), Springer, HEP (2010) 1--26.

\bibitem{ALM13}
R. Aguilar-Sanchez, E. D. Leonel, and J. A. Mendez-Bermudez,
Dynamical properties of a dissipative discontinuous map: A scaling investigation,
Phys. Lett. A {\bf 377}, 3216 (2013).

\bibitem{MOL16}
J. A. Mendez-Bermudez, J. A. deOliveira, and E. D. Leonel,
Analytical description of critical dynamics for two--dimensional dissipative nonlinear maps,
Phys. Lett. A {\bf 380}, 1959 (2016).

\bibitem{MASL24}
J. A. Mendez-Bermudez, R. Aguilar-Sanchez, J. M. Sigarreta, and E. D. Leonel,
Scaling properties of the action in the Riemann-Liouville fractional standard map,
Phys. Rev. E {\bf 109}, 034214 (2024).


\end{thebibliography}
\end{document}